# CrossBench: Generalized Crosstalk Benchmark Generation for Quantum Computers

Jaden Hawley
*MU Quantum Innovation Center*
*Electrical Engineering and Computer Science*
*University of Missouri - Columbia*
Columbia, MO, USA
jrh22r@umsystem.edu

Chi-Ren Shyu
*MU Quantum Innovation Center*
*Electrical Engineering and Computer Science*
*University of Missouri - Columbia*
Columbia, MO, USA
shyuc@missouri.edu

*Abstract*— **As quantum computers continue to increase in size and topological complexity, benchmarking crosstalk becomes more complex and resource-intensive. This limits the ability to obtain relevant crosstalk metrics for applications such as error mitigation, quantum computer security, and circuit transpilation. These applications benefit from accessible metrics on how each gate contributes to crosstalk. However, crosstalk metrics are rarely provided by quantum computer providers and can be expensive to obtain on modern large NISQ devices. In this work, we propose CrossBench, a customizable system that generates crosstalk benchmarks for finitely large NISQ devices with arbitrary topologies. CrossBench uses a custom graph labeling algorithm to generate crosstalk benchmarks for a given quantum topology and gate set. These benchmarks can be used to estimate the average contribution of each gate's crosstalk to qubit error rates. We evaluate the effectiveness of CrossBench by generating and running benchmarks on multiple IBM quantum computers with different topologies. Our results show that CrossBench can identify gates that introduce significant crosstalk across all tested devices, with strong statistical significance ($p < 0.05$). These promising results show that CrossBench can give simple and accessible crosstalk benchmarks for modern NISQ systems.**

*Keywords—Quantum Computer, Crosstalk, Benchmark*

## I. INTRODUCTION

Crosstalk errors in quantum computers occur when operations on one qubit unintentionally affect other qubits [1]. In superconducting machines, one source of crosstalk is an always-on ZZ interaction between coupled qubits [2], [3]. However, crosstalk can also arise from other types of interactions, such as ZX and XY couplings [4], [5]. Importantly, crosstalk can occur on qubits that are not directly connected [6] and can also appear in other hardware platforms, such as neutral atom systems [7]. Because of these varied sources, we define crosstalk broadly as an error in which gates on one qubit unintentionally influence the state of another qubit [1]. Crosstalk is relevant to many areas in quantum computing beyond error mitigation and characterization, such as security [8], [9], [10], [11], [12], [13] and multitenancy [7]. For these applications, it is important to have clear and accessible crosstalk metrics. However, such metrics are often not provided by quantum hardware vendors. To obtain crosstalk metrics, a common approach is to use crosstalk characterization and benchmarking techniques. A crosstalk benchmark can serve multiple purposes, such as detecting weather crosstalk is present, identifying which gates generate more crosstalk, and determining the strength of crosstalk. There is a considerable body of prior work characterizing and benchmarking crosstalk in quantum computing systems, providing a solid foundation that we build on and extend in this study.

Two widely used techniques are *Idle Tomography (IDT)* [1], [8], [14], and *Simultaneous Randomized Benchmarking (SRB)* [15]. IDT designates a *driver* qubit on which gates are executed, while the remaining qubits serve as idle *spectators* and are measured after the driver qubit completes its operations. For two-qubit gates, pairs of driver qubits are used. This process is repeated on all qubits or qubit pairs to accurately model detailed crosstalk behavior. SRB involves running Randomized Benchmarking (RB) circuits on each qubit, similarly to IDT. However, the measurement gate is applied to the driver qubits instead of the idle qubits. The next step is to perform two RB circuits simultaneously and compare them to the individual RB circuit results to determine the crosstalk effect of different gates. While SRB and IDT are effective, they require exhaustive testing of each qubit or coupling, which becomes increasingly costly as system size and connectivity grow, limiting scalability [16]. These methods remain costly on large scale-systems that surpass 100 qubits.

Several efforts have aimed to make crosstalk benchmarking more scalable. These include cycle benchmarking, a general noise characterization approach whose cost does not scale with system size [16]; subsystem-based designs for quantum computers, such as those used in fault-tolerant architectures [17]; and the use of native gates, rather than Clifford gates, in RB [18]. In this work, we focus on a simple, effective benchmark that targets crosstalk on NISQ systems and supports automatic benchmark generation. We propose *CrossBench*, a customizable, scalable crosstalk benchmark generator. CrossBench provides aggregate metrics that capture overall device behavior and shows how different gates on a quantum computer affect each other through crosstalk. CrossBench is designed to maintain constant quantum compute cost as system size grows and is highly customizable. This flexibility allows for error tolerances and sensitivity to be adjusted, which can allow users to adjust for quantum systems that have weaker crosstalk.

Emerging attack strategies such as QubitVise exploit crosstalk more effectively by inducing interference from multiple qubits simultaneously [19]. In contrast, existing crosstalk modeling and benchmarking approaches, such as IDT, consider only a single source of crosstalk. CrossBench enables the benchmarking of multiple sources simultaneously. Crosstalk remains a concern across multiple areas of quantum computing, including quantum cybersecurity [8], [9], [10], [11], [12], [13], error mitigation and characterization [16], [17], [18], [20], and multitenancy [7], [8], [9], [10], [11]. There is a clear need for scalable and accessible methods that specifically measure crosstalk. CrossBench offers a unique, scalable, and customizable benchmark as an option to address this need.

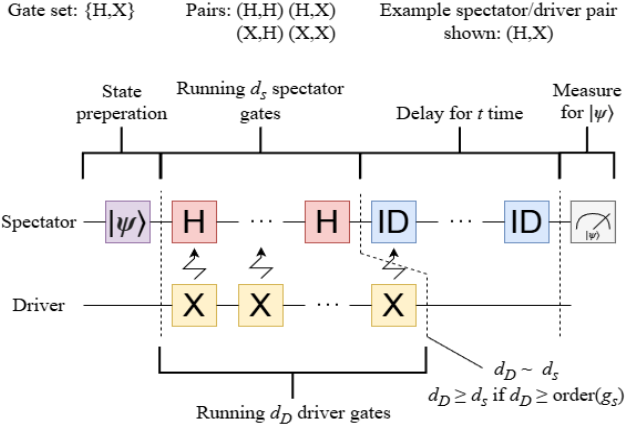

Fig. 1. Example of spectator and driver qubits generated from spectator and driver pair (H,X). The driver generates crosstalk and the spectator attempts to observe it.

## II. METHODS

CrossBench generates a set of benchmark circuits that measure the crosstalk induced by each provided gate on a system. It specifically estimates the change in error experienced by a qubit when other qubits are executing a given gate. Following prior work [1], we will utilize *driver* and *spectator* qubits. Spectator qubits are measured at the end of the benchmark and are used to detect crosstalk effects. Driver qubits are used to generate crosstalk by repeatedly applying a given gate and are not measured. Figure 1 illustrates this. The number of repetitions for the driver qubit, $d_D$, is determined by a user-provided gate error rate and error threshold, as defined by Eq. 1. Let $E$ denote the estimated error probability of the maximum error rate among native gates on the quantum computer and let $\delta$ denote the acceptable gate error threshold. This depth is used for both driver qubits and as a reference for determining spectator depth. It is chosen such that the cumulative error from applying $d_D$ gates remains approximately bounded by $\delta$.

$$d_D = 10^{-\lfloor \log E - 2 \rfloor} * \delta \quad (1)$$

To generate benchmarks, CrossBench requires inputs describing the target quantum computer. The user provides the native gates to evaluate, along with their durations and *orders*. Gate order refers to the number of repetitions required for the gate to be equivalent to identity to a global phase, shown in Eq. 2. Devices that support arbitrary or parameterized gates require a finite subset of gates to be specified.

$$\text{order}(U) = n, \quad U^n = e^{i\theta} I \quad (2)$$

The topology is specified as a list of edges, along with a flag indicating whether it is directed or undirected. For directed topologies, multi-qubit gates are constrained to follow the edge orientation, where the ordering of vertices in each edge defines the direction. For undirected edges, both ordered pairs for each edge are included. Undirected topologies are agnostic to gate orientation or only support symmetric gates, and each edge is listed only once. CrossBench constructs all ordered pairs of the given gates. Each pair defines a benchmark configuration: one gate is applied to spectator qubits, and the other is applied to driver qubits. CrossBench generates one benchmark for each gate pair. Figure 1 shows examples of spectator and driver qubits derived from an ordered pair.

### A. Benchmark Generation

The benchmark generation process constructs a graph representation of the device topology and then applies four passes to assign driver and spectator qubits. Algorithm 1 shows the overall process of CrossBench and can be referenced through this section. CrossBench represents the quantum computer topology as a directed graph in which vertices correspond to qubits and edges correspond to couplings. The direction of an edge is determined by how entangling gates are oriented on the coupling they represent. If a quantum coupling allows for any orientation, then it is treated as an undirected edge.

CrossBench then places spectator and driver qubits randomly on the quantum computer. The placement is constrained such that all spectator qubits have at least one neighboring driver qubit, and the numbers of driver and spectator qubits are balanced as much as possible. This is achieved by setting two thresholds: a driver threshold and a spectator threshold. These act as soft minimum targets for the number of qubits assigned to each role. These may be set by the user. Higher driver thresholds can result in increased sensitivity to crosstalk, while higher spectator thresholds result in a larger qubit sample size.

There are four passes of placing driver qubits and spectator qubits based on these thresholds. If any pass can no longer assign valid qubits, then it terminates and proceeds to the next pass. Pass 1 places driver qubits until the driver threshold is reached, and Pass 2 places spectator qubits until the spectator threshold is reached. Pass 3 places driver qubits without enforcing the threshold, and Pass 4 places spectator qubits without enforcing the threshold.

The first two passes place driver and spectator qubits up to the set threshold. The last two passes are to fill any remaining space and optimize sample size and sensitivity. The user may choose to skip Pass 3 and Pass 4 to strictly follow the threshold.

| Algorithm 1: CrossBench |   |
|---|---|
| 1: | Calculate driver depth using Eq. 1 |
| 2: | Generate Ordered Gate Pairs |
| 3: | For each (Spectator, Driver) in pairs |
| 4: |    Initilize Circuit |
| 5: |    Pass 1: AssignDriver() Until Threshold |
| 6: |    Pass 2: AssignSpectator() Until Threshold |
| 7: |    Pass 3: AssignDrivers() |
| 8: |    Pass 4: AssignSpectators() |
| 9: |    AddMeasurementGates(Spectators) |
| 10: |    Add Circuit to benchmark list |
| 11: | End loop |

Algorithm 2 shows pseudocode for pass 1. Pass 2 functions the same but by swapping driver and spectator. Pass 3 and 4 function the same as Pass 1 and 2 but without the threshold.

During assignment, each selected vertex is labeled as either a driver or spectator. To ensure that every spectator has at least one neighboring driver and every driver has at least one neighboring spectator, the placement algorithm checks whether a newly placed qubit has at least one neighbor of the opposite type. If no such neighbor exists, then it will place a new neighbor randomly. This is a requirement for placing either qubit type.

*B. Spectator Qubit Construction*

Each spectator qubit is first prepared in one of six states, chosen uniformly at random: $|0\rangle, |1\rangle, |+\rangle, |-\rangle, |i\rangle,$ or $|-i\rangle$, or from a user-specified set of states. A sequence of gates is then applied according to the selected benchmark gate pair. Let $g_s$ denote the spectator gate and $d_D$ denote the driver depth defined earlier. Spectator depth $d_s$ is given by Eq. 3 and Eq. 4.

$$d_s = \text{order}(g_s) * \left\lfloor \frac{d_D}{\text{order}(g_s)} \right\rfloor \quad (3)$$

$$\text{if order}(g_s) > d_D, \text{then } d_s = \text{order}(g_s) \quad (4)$$

This construction ensures that the applied gate sequence is equivalent to the identity operation, while keeping the total number of applied gates below $d_D$. This results in the expected error contribution from spectator gates remaining bounded by the user-provided threshold except as described in Eq. 4. Figure 1 illustrates the timing by showing a circuit diagram of how driver and spectator qubits are constructed.

After these gates are applied, the spectator qubit is delayed before measurement. This delay ensures the driver qubits have enough time to complete their gates, and all benchmarks within a set experience comparable decoherence effects. Let $g_{max}$ denote the gate with the longest duration among the native gate set. The required delay time $t$ is defined by Eq. 5. This aligns the total execution time of the spectator qubit with the longest possible driver execution time. After the delay, the initial prepared state is reversed, and measurement gates are applied.

$$t = \text{duration}(g_{max})d_D - \text{duration}(g_s)d_s \quad (5)$$

When a driver qubit is placed, the driver gate $g_D$ is applied repeatedly for $d_D$ layers. Driver qubits are not initialized in a specific state and are not measured, as their sole purpose is to generate crosstalk on neighboring spectator qubits.

*C. Graph Labeling*

When either a driver qubit or a spectator qubit is assigned, CrossBench must find a valid empty location on the device topology. A valid location has at least one neighboring qubit of the opposite role or it can be assigned one.

To achieve this, CrossBench iterates through a shuffled list of all the qubits. For each qubit, it checks whether it is valid for placement using a graph-labeling algorithm using that qubit as a root node.

*D. Driver Assignment*

To assign a driver qubit, the algorithm begins at a candidate qubit and examines its neighbors. If there exists a neighboring spectator qubit, the placement can proceed. If no neighboring spectator qubit exists, CrossBench searches for unassigned neighbors that can be assigned as spectator qubits. If such a neighbor is found, it is temporarily marked to prevent conflicts during the search. If no neighbor is found or assigned, then the placement is invalid, and CrossBench continues to the next candidate qubit. During placement, CrossBench maintains two flags marking if a neighbor was found or assigned.

Pseudocode is shown in Algorithm 3 for CrossBench qubit placement. CrossBench then recursively explores all neighbors. At each recursive step it propagates information about the state. This information includes the number of remaining driver qubits to assign to allow for placement of the driver gate, if a spectator neighbor has already been found or assigned, and the traversal direction. As CrossBench traverses the graph, it temporarily labels visited nodes to prevent conflicts while searching. If a spectator neighbor is found after candidate spectator qubits were previously identified, the temporary mark on the candidate qubit is removed, and the state is updated to reflect that a spectator neighbor has already been found.

Once it reaches the last driver qubit that needs to be assigned, it checks the neighboring spectator qubit constraint. If no neighboring spectator qubit was found or assigned; the placement is invalid, and the algorithm backtracks to explore alternative placements. If a neighboring spectator qubit was found or can be assigned, the placement is valid and begins a return chain to mark qubits and generate a qubit list. On valid placement, the algorithm returns an ordered list of new driver and spectator qubits and metadata indicating the required orientation for multi-qubit gates. If no spectator qubit was found or placed, the algorithm returns that the current driver placement is invalid, allowing previous iterations to search for other placements.

| Algorithm 2: CrossBench Pass 1 |
| --- |
| 1:   Loop over Shuffled Qubits |
| 2:       TryAssignDriver(Qubit, Gate Order) |
| 3:       If success |
| 4:           Place Driver and Spectator gates |
| 5:           If Driver count ≥ Threshold |
| 6:               Break |
| 7:       Else |
| 8:           Continue |

| Algorithm 3: CrossBench Driver Placement |
| --- |
| 1:   If Gate Order = 1 |
| 2:       If current qubit is valid |
| 3:           If Neighboring Spectator is placed or found |
| 4:               Return list of new drivers and spectators |
| 5:   Else |
| 6:       If current qubit is valid |
| 7:           Assign current qubit as driver |
| 8:           If No Neighboring Spectator previously found |
| 9:               Search neighbors for spectator |
| 10:              If no neighboring spectator previously placed |
| 11:                  Try placing neighboring spectator |
| 12:          Loop over Shuffled Neighbors |
| 13:              TryAssignDriver(Neighbor, Gate Order - 1) |
| 14:              If Success |
| 15:                  If neighboring spectator found |
| 16:                      Unassign any placed spectator |
| 17:                  Return list of new drivers and spectators |
| 18:          Unassign new Drivers and Spectators |
| 19:   Return Invalid Placement |

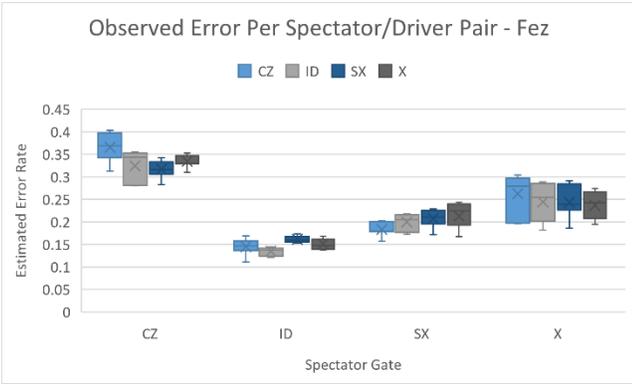

Fig. 2. Results for each driver spectator gate pair on IBM Fez. The colors represent the Driver gate, while each cluster of box and whisker charts represents a spectator gate.

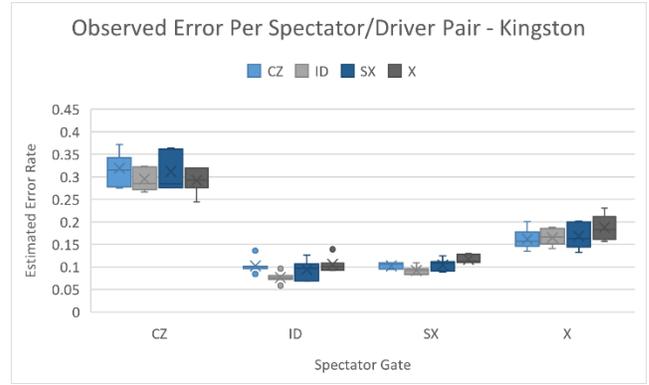

Fig. 3. Results for each driver spectator gate pair on IBM Kingston. The colors represent the Driver gate, while each cluster of box and whisker charts represents a spectator gate.

*E. Benchmark Execution*

The output of CrossBench is a list of Qiskit circuits, along with metadata describing each circuit. Each circuit corresponds to a specific ordered pair of driver and spectator gates. If the native gate set is utilized, minimal transpilation is required. The returned metadata contains a spectator count, which determines the length of the output bitstrings. After execution, each circuit produces results in the form of binary strings, where each bit corresponds to a measured spectator qubit: a "0" bit indicates no error detected, and a "1" bit indicates an error detected.

The detected error can arise from multiple sources including State Preparation and Measurement (SPAM) errors, crosstalk, decoherence, and gate errors. By aggregating these results over multiple shots, one can compute metrics such as the average error rate for each benchmark circuit. Since each circuit corresponds to a specific gate pair, these metrics quantify the impact of one gate's crosstalk on another.

### III. RESULTS AND DISSCUSSION

*A. Results*

We used CrossBench to generate and execute benchmarks on multiple IBM quantum computers. These systems include two 156-qubit Heron processors, IBM Fez and IBM Kingston, and one 120-qubit Nighthawk processor, IBM Miami. Each benchmark was run with 10,000 shots, gate error threshold $\delta = 0.1$, and driver/spectator qubit thresholds set to evenly balance each role's gate counts. Seven sets of benchmarks were generated for each device to capture the variability across runs. The gates evaluated for driver and spectator qubits are the X, SX, CZ, and ID gates. These gates were chosen because they are the available non-parameterized, non-virtual gates on all the tested machines [21][22].

These benchmarks produce binary string outputs, where a "1" bit indicates that an error is detected on a spectator qubit. An estimated error rate can be derived by calculating the total number of "1" measurements and dividing it by the total number of measurements. The total number of measurements can be calculated by multiplying the shot count and spectator qubit count for a given benchmark.

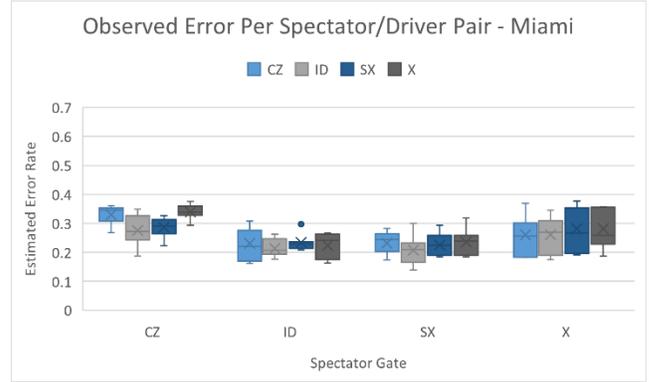

Fig. 4. Results for each driver spectator gate pair on IBM Miami. The colors represent the Driver gate, while each cluster of box and whisker charts represents a spectator gate.

This estimated error rate is representative of multiple noise sources including crosstalk, decoherence, gate error, and SPAM errors. However, when comparing benchmarks that use the same spectator gates, the primary difference in error rates is what driver gate was used and the random layout of the benchmark. Therefore, consistent differences in estimated error rates can be attributed to driver gate induced crosstalk.

Figure 2 shows box-and-whisker plots of the results from all seven runs for IBM Fez. We observe a noticeable difference in error on ID and CZ spectator gates. Using a t-test, we find ID spectators experience more error with SX drivers than ID drivers ($p < 0.01$). We also find CZ spectators experience more error with CZ drivers than SX drivers ($p = 0.021$). Figure 3 shows the results for IBM Kingston. For this system, ID spectators experience higher error with X drivers than ID drivers ($p < 0.01$) and experience more error with CZ drivers than ID drivers ($p = 0.046$). Additionally, SX spectators experience more error with X drivers than with ID drivers ($p < 0.01$). For the Nighthawk system, Figure 4 shows the results for IBM Miami. We find CZ spectators experience higher error with X drivers than ID drivers ($p = 0.021$) and SX drivers ($p < 0.01$).

TABLE I.    FEZ ERROR PROBABILITY PER GATE PAIR

| IBM Fez | | Driver Gate | | | |
|---|---|---|---|---|---|
| | | CZ | ID | SX | X |
| Spectator Gate | CZ | 0.365 | 0.324 | 0.317 | 0.334 |
| | ID | 0.145 | 0.135 | 0.161 | 0.151 |
| | SX | 0.182 | 0.200 | 0.207 | 0.212 |
| | X | 0.261 | 0.244 | 0.244 | 0.236 |

Table i. Error probabilities for IBM Fez. Averaged over seven benchmark runs. The spectator qubit gate is what gate the qubit we measure runs. The neighbor gates are what gate we are testing the crosstalk of.

TABLE II.    KINGSTON ERROR PROBABILITY PER GATE PAIR.

| IBM Kingston | | Driver Gate | | | |
|---|---|---|---|---|---|
| | | CZ | ID | SX | X |
| Spectator Gate | CZ | 0.319 | 0.294 | 0.310 | 0.292 |
| | ID | 0.102 | 0.076 | 0.094 | 0.105 |
| | SX | 0.102 | 0.092 | 0.104 | 0.117 |
| | X | 0.162 | 0.164 | 0.169 | 0.187 |

Table ii. Error probabilities for IBM Kingston. Averaged over seven benchmark runs. The spectator qubit gate is what gate the qubit we measure runs. The neighbor gates are what gate we are testing the crosstalk of.

TABLE III.    MIAMI ERROR PROBABILITY PER GATE PAIR.

| IBM Miami | | Driver Gate | | | |
|---|---|---|---|---|---|
| | | CZ | ID | SX | X |
| Spectator Gate | CZ | 0.332 | 0.275 | 0.287 | 0.339 |
| | ID | 0.230 | 0.212 | 0.234 | 0.223 |
| | SX | 0.317 | 0.207 | 0.225 | 0.235 |
| | X | 0.334 | 0.260 | 0.282 | 0.282 |

Table iii. Error probabilities for IBM Miami. Averaged over seven benchmark runs. The spectator qubit gate is what gate the qubit we measure runs. The neighbor gates are what gate we are testing the crosstalk of.

Table 1 contains the averaged estimated error rate reported across seven benchmarks for each spectator and driver qubit pair for IBM Fez. Table 2 shows IBM Kingston's results, and Table 3 shows IBM Miami's results. Differences in observed error probability between driver gates for the same spectator gates can be attributed to crosstalk. An example of this is that the CZ driver consistently produces higher error rates than an ID driver on every spectator gate on IBM Miami.

An average error rate across all spectator gates can be calculated for each driver gate. This provides insight into which driver gates produce more crosstalk to an unknown or mixed workload. Because individual benchmarks may show variability, repeating them increases the stability of the results. We next averaged the results of all seven runs and show them in the following figures with error bars. We found that seven runs provide a good balance between statistical stability and quantum compute cost for the Heron processors.

Figure 5 shows averaged CrossBench results across all spectator gates for IBM Miami. Figure 6 shows results for IBM Kingston and Figure 7 shows results for IBM Fez. The lowest

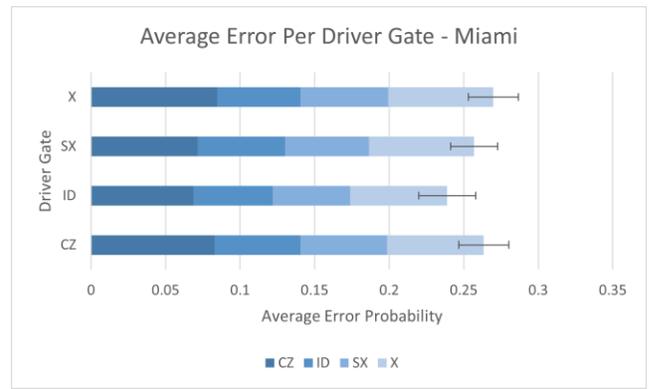

Fig. 5. The average error rate per driver gate observed on IBM Miami. Each bar is sectioned by color to show what portion of the error rate was observed with a given spectator qubit gate. Standard error margins are shown.

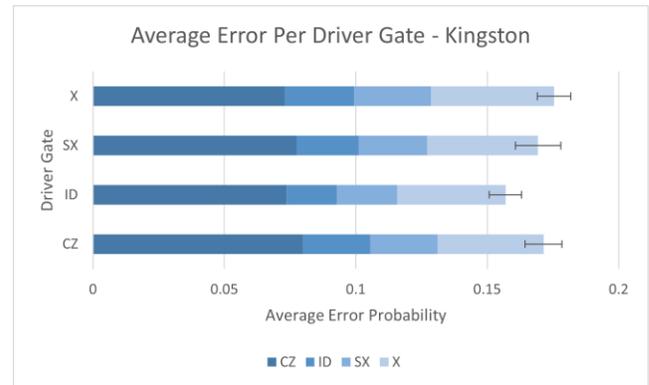

Fig. 6. The average error rate per driver gate observed on IBM Miami. Each bar is sectioned by color to show what portion of the error rate was observed with a given spectator qubit gate. Standard error margins are shown.

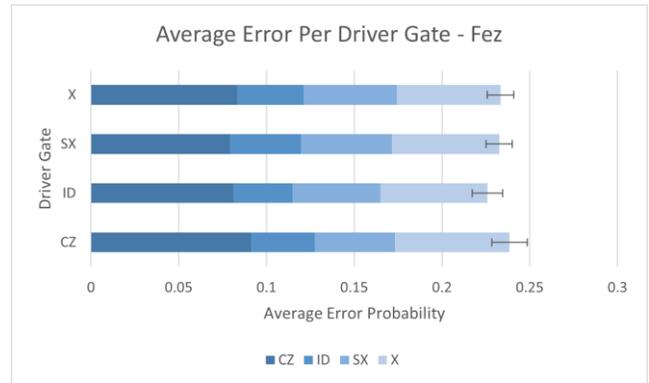

Fig. 7. The average error rate per driver gate observed on IBM Miami. Each bar is sectioned by color to show what portion of the error rate was observed with a given spectator qubit gate. Standard error margins are shown.

error rates were consistently observed across all devices with ID drivers.

*B. Discussion*

These results show that benchmarks generated by CrossBench can extract useful information about the crosstalk of large NISQ systems. Using identity gates as a baseline, the crosstalk contribution of a given driver gate can be estimated by subtracting the corresponding ID-driver results. This provides a simple and interpretable measure of relative crosstalk strength. However, this approach assumes that ID gates introduce negligible additional error. On devices where ID gates are implemented via pulses, they may still contribute to noise. On such devices, a more accurate baseline can be obtained by running a separate control batch without any gates on the driver qubits. An alternative option is to use the driver gate with the lowest observed spectator error as a baseline to get a lower-bound estimate of crosstalk.

Utilizing multiple drivers provides valuable insights to observing crosstalk effects from combined and distant neighbors seen in modern crosstalk attacks [6], [19] and can be used to detect crosstalk in machines with weaker crosstalk.

The benchmarks exhibit some run-to-run variability. This variability arises from the random nature of CrossBench, quantum noise, and changes in hardware performance over time. Repeating the benchmarks multiple times reduces this uncertainty. Different quantum computers exhibit different levels of crosstalk and variability. This indicates that more repetitions may be needed to obtain stable estimates. As a result, there is no single fixed number of runs that guarantees accuracy across all systems.

CrossBench was evaluated on a variety of topologies and quantum machines, including Heron and Nighthawk processors. These experiments demonstrate that CrossBench is applicable across diverse architectures. More generally, this method is compatible with any quantum computer that has a topology that can be represented by a finite directed graph.

Similar error behaviors to crosstalk have been observed outside of superconducting machines as well, such as neutral atom machines [7]. Because CrossBench relies only on topology and gate definitions, it can be extended to these platforms to study errors that arise when neighboring qubits operate simultaneously. This suggests that CrossBench provides a broadly applicable framework for characterizing interaction-induced errors across different quantum computing technologies. Overall, these results show that CrossBench provides a practical and scalable method for isolating and comparing crosstalk effects across large quantum systems, while remaining flexible across hardware architectures.

## IV. CONCLUSION

In this work, we presented CrossBench, a tool for generating scalable crosstalk benchmarks on quantum computers. Its effectiveness was demonstrated by applying it to multiple IBM systems, where it successfully detected gate-dependent-variations caused by crosstalk. It identified gates that induce crosstalk on every NISQ system benchmarked in this paper with strong statistical significance ($p < 0.05$). CrossBench can produce benchmarks for arbitrary directed graph topologies and finite unitary gate sets, as demonstrated by its successful execution on multiple QPU architectures. This makes CrossBench an effective tool for measuring crosstalk on large machines where SRB and IDT are prohibitively expensive. CrossBench is a flexible tool that can generate benchmarks that are highly customizable in terms of error tolerances, sensitivity, and sampling.

CrossBench shows promise as a useful tool for obtaining inexpensive crosstalk metrics with high customizability. Its ability to have multiple driver qubits is relevant for defending against modern quantum computer security threats such as QubitHammer [6], where it could assist in crosstalk-based-malware detection.

### A. Limitations

Early versions of CrossBench were successfully run on IBM Eagle r3 and IBM Heron r1 processors, which added additional variation to tested gates and topologies. However, these systems were retired before the final results could be obtained and are therefore not reported in this work. To compensate for the loss of Eagle results, we included the newer Nighthawk processor, IBM Miami, to introduce additional variation in topologies. For IBM Miami, we found that the same jobs required up to twelve times more quantum compute time as other IBM machines, which limited how many benchmarks sets we could run and our ability to fine-tune parameters for the IBM Miami [23]. To address this, we used the lower-cost Heron QPUs to estimate parameters. The increase in compute time is temporary and we plan to re-run CrossBench on IBM Miami in the future once performance improves.

### B. Future Work

We plan to make CrossBench available for broader use after further improvements and documentation are completed. Planned improvements include allowing users to specify padding between spectator and driver qubits. This will allow for benchmarking the effectiveness of different circuit paddings on NISQ systems. We will continue to benchmark all the systems made available to us. In future work, we also plan to use this tool to develop methods for crosstalk-based-malware detection and mitigation, as well as multitenant scheduling.

ACKNOWLEDGMENT

We would like to thank the Mizzou Quantum Innovation Center, College of Engineering, and College of Arts & Sciences for supporting this research. This research was supported by the National Science Foundation under the grant number DEG-1946619.